\documentclass[conference]{IEEEtran}
\IEEEoverridecommandlockouts
\pdfoutput=1
\usepackage{amsmath,amssymb,amsfonts}
\usepackage{graphicx}
\usepackage[numbers,sort&compress]{natbib}
\usepackage{graphicx}%
\usepackage{multirow}%
\usepackage{amsmath,amssymb,amsfonts}%
\usepackage{amsthm}%
\usepackage{mathrsfs}%
\usepackage{array}
\usepackage{xcolor}%
\usepackage{textcomp}%
\usepackage{manyfoot}%
\usepackage{booktabs}%
\usepackage{algorithm}%
\usepackage{algorithmicx}%
\usepackage{algpseudocode}%
\usepackage{listings}%
\usepackage{subfigure} 
\usepackage{upgreek} 
\usepackage{tabularx}
\usepackage{amsmath,amsthm,amssymb,amsfonts}
\usepackage{siunitx}
\usepackage{threeparttable}
\usepackage{multirow}
\usepackage{makecell}  
\usepackage{textcomp}
\usepackage{xcolor}

\def\BibTeX{{\rm B\kern-.05em{\sc i\kern-.025em b}\kern-.08em
    T\kern-.1667em\lower.7ex\hbox{E}\kern-.125emX}}
\begin{document}

\title{Introducing the Brand New QiandaoEar22 Dataset for Specific Ship Identification Using Ship-Radiated Noise\\

\thanks{This work was funded jointly by Natural Science Foundation of Shanghai (Grant No. 22ZR1475700), Youth Innovation Promotion Association CAS (Grant No. 2021022), and Young Talent Cultivation Program of Shanghai Branch of CAS.}
}

\author{\IEEEauthorblockN{1\textsuperscript{st} Xiaoyang Du}
\IEEEauthorblockA{
\textit{Shanghai Acoustics Laboratory,}\\
\textit{Chinese Academy of Sciences}\\
\textit{University of Chinese Academy of Sciences}\\
Shanghai, China \\
duxiaoyang22@mails.ucas.ac.cn}
\and
\IEEEauthorblockN{2\textsuperscript{nd} Feng Hong}
\IEEEauthorblockA{
\textit{Shanghai Acoustics Laboratory,}\\
\textit{Chinese Academy of Sciences}\\
Shanghai, China \\
hongfeng@mail.ioa.ac.cn}
}
\maketitle    
\begin{abstract}
Target identification of ship-radiated noise is a crucial area in underwater target recognition. However, there is currently a lack of multi-target ship datasets that accurately represent real-world underwater acoustic conditions. To ntackle this issue, we release QiandaoEar22 \textemdash an underwater acoustic multi-target dataset, which can be download on https://ieee-dataport.org/documents/qiandaoear22. This dataset encompasses 9 hours and 28 minutes of real-world ship-radiated noise data and 21 hours and 58 minutes of background noise data. We demonstrate the availability of QiandaoEar22 by conducting an experiment of identifying specific ship from the multiple targets. Taking different features as the input and six deep learning networks as classifier, we evaluate the baseline performance of different methods. The experimental results reveal that identifying the specific target of UUV from others can achieve the optimal recognition accuracy of 97.78\%, and we find using spectrum and MFCC as feature inputs and DenseNet as the classifier can achieve better recognition performance. Our work not only establishes a benchmark for the dataset but helps the further development of innovative methods for the tasks of underwater acoustic target detection (UATD) and underwater acoustic target recognition(UATR).
\end{abstract}


\section{Introduction}
The importance of underwater acoustic target has been emphasized since American inventor Fessenden's creation of sonar in 1913. However, publicly available datasets for underwater acoustic target research are scarce. Some small scale datasets have been created, including those from cargo ships, navy vessels, and personal watercraft. While some recent datasets like ShipsEar \cite{santos2016shipsear} and Deepship \cite{irfan2021deepship} are available, they primarily consist of single target ship-radiated noise data in simple environments, lacking the complexity of multiple target scenarios which often encounter in real-world conditions.  

In this paper, an underwater acoustic multi-target dataset named QiandaoEar22, is constructed through the experiment conducted in Qiandao Lake in 2022. Then, an experimental task is demonstrated to identify the specific ship from the multiple targets, which could be a reference when using this dataset for other potential purposes.

\section{The QiandaoEar22 dataset}

The data collection was performed from June 24 to 28, 2022, at the Xin' anjiang Experimental Site of the Institute of Acoustics, Chinese Academy of Sciences (CAS), situated in the Qiandaohu town, Zhejiang, China. The coordinates of our sampling locations were 118.974532\textdegree E, 29.557795\textdegree N, and 118.947097\textdegree E, 29.548708\textdegree N, respectively. During the experiments, a DigitalHyd SR-1 hydrophone was anchored at the center of the deployment area using a counterweight and a buoy. The water depth was 30 to 50 meters, while the hydrophone was positioned at a depth of 10 to 15 meters. The experimental equipment was deployed near the channel to capture a broader range of radiated noise from the ships.

\begin{figure}[htbp]  
	\centering
	\includegraphics[width=0.5\textwidth]{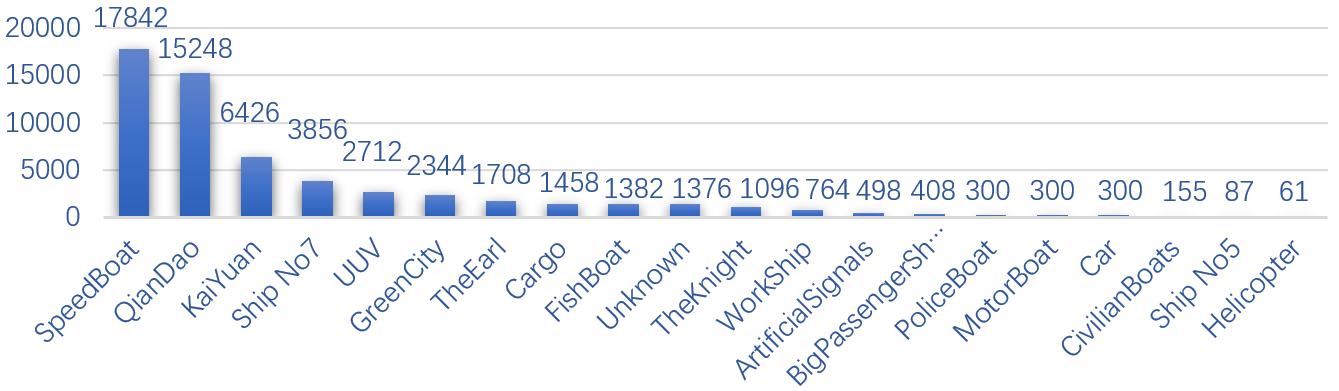}
	\caption{Duration of the recorded data for different targets}
	\label{distribution}
\end{figure}

The recorded targets include large tour boats, cargo ships, sanitation vessels, small boats, research ships, and other targets such as helicopters and cars which noise were produced in the air but spread into underwater. The section of each audio segment are recorded according to the spectral characteristics. We label each data based on auditory sensation, distance, and their types. Among these tags, the auditory sensation is a subjective indicator which means the sound intensity of what the human ears hear, which can be strong,middle or weak. After labeling, the segmented data has the length of 3 seconds. 

The ship-radiated noise dataset consists of 10,611 records in WAV format with a total of 143 recordings containing single or multiple ship targets. As shown in Fig. \ref{distribution}, our dataset contains 20 categories of targets. Additionally, in order to ensure the integrity of the constructed dataset, we collect some background noise data during nighttime and idle periods. Ultimately, our dataset obtains a total of 9 hours and 28 minutes of real-world ship-radiated noise data and 21 hours and 58 minutes background data, which can be download on https://ieee-dataport.org/documents/qiandaoear22. 

\section{Experimental task: Identifying a specific type of ship signals from multiple targets}
\begin{figure}[t!]  
	
	\begin{center}
		\centering
		\subfigure[SpeedBoat]{
			\includegraphics[width=0.13\textwidth,height=0.5in]{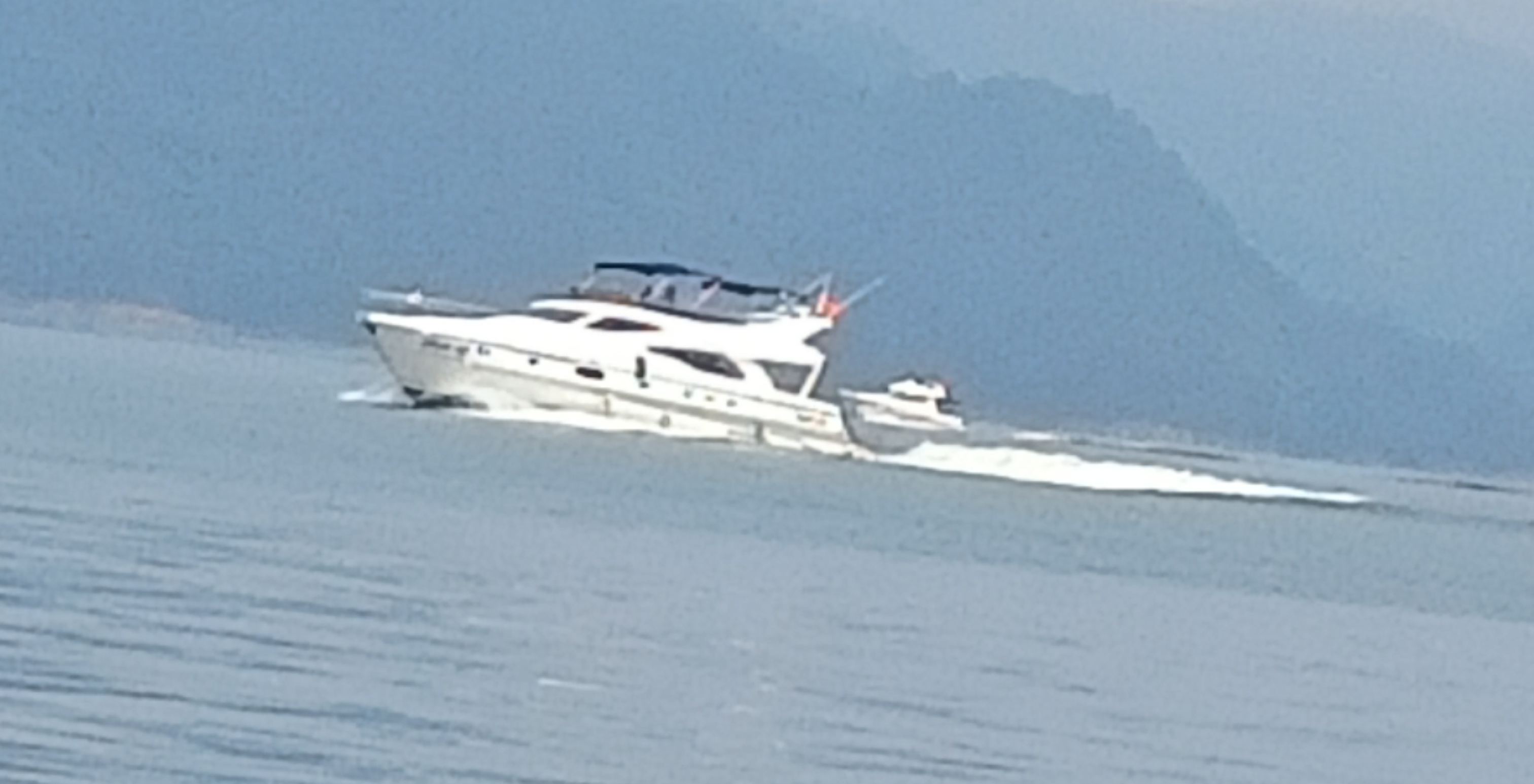}
		}
		\subfigure[KaiYuan]{
			\includegraphics[width=0.13\textwidth,height=0.5in]{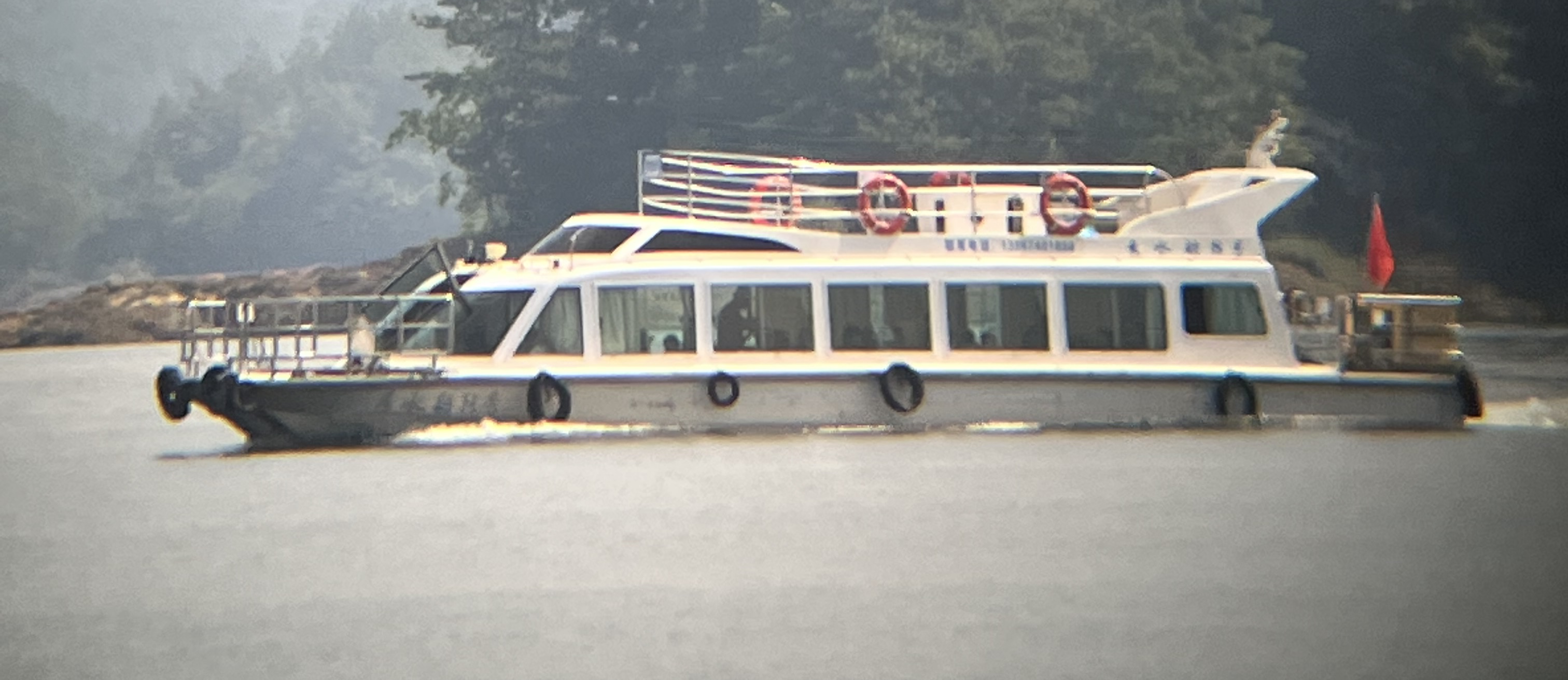}
		}
		\subfigure[UUV]{
			\includegraphics[width=0.13\textwidth,height=0.5in]{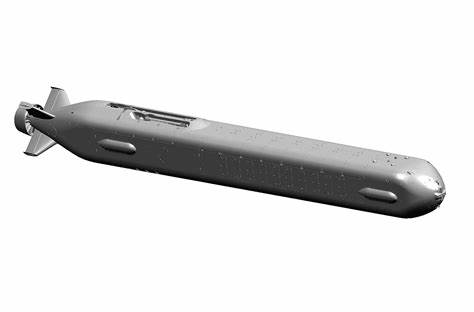}
		}
		\caption{Target ships}
		\label{boat}
	\end{center}
\end{figure}

To demonstrate the use of the dataset, we design an experimental task to identify a specific type of ship signals from the multi-target ship-radiated noise signals. The ship types can be speedboat, KaiYuan, and UUV, which is shown in Fig. \ref{boat}. We take different features as the input and several deep learning networks as classifier to evaluate the baseline performance of different methods. 

\begin{figure}  
	\centering
	\includegraphics[width=0.5\textwidth]{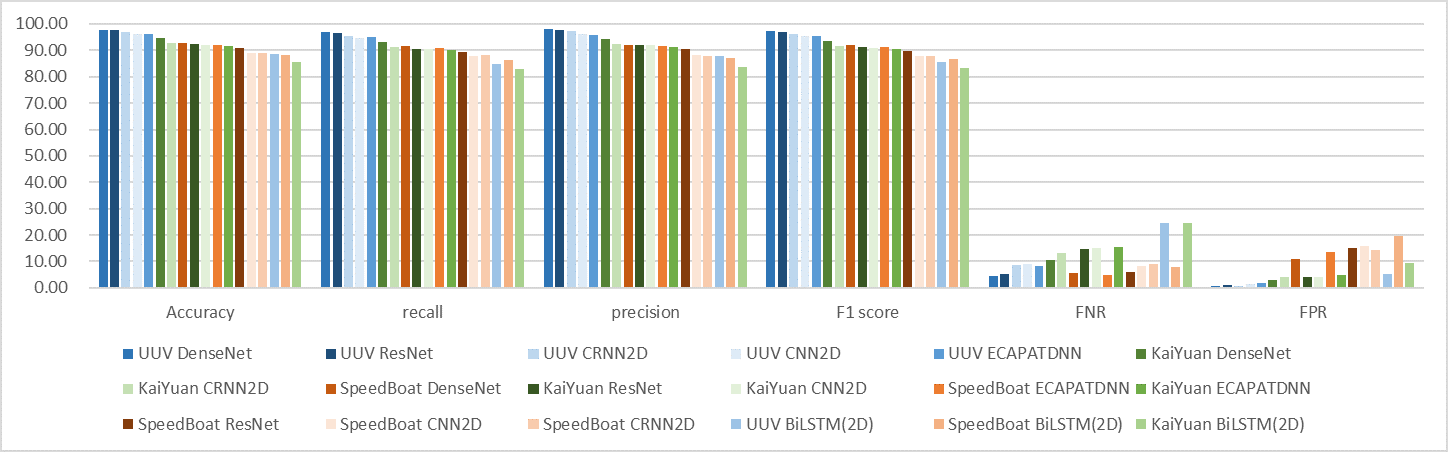}
	\caption{Average experimental results of deep learning methods for 2D feature classification}
	\label{2dresultfig}
\end{figure}

To compare the performance of different networks, we calculate the average performance for different 2D features using the same network, shown in Fig. \ref{2dresultfig}. Taking the UUV experiment as an example, DenseNet \cite{yu2021additive} performs best, followed by ResNet, CRNN, CNN, ECAPA-TDNN, and BiLSTM. The best average recognition accuracy can reach 97.78\%, with only 4.34\% of the data containing UUV targets being falsely recognized. To summarize, empirically, the DenseNet can be an exceptional choice when selecting a classifier to identify a specific type of ship signals from multiple targets using 2D features.

Then, we average the results obtained for each feature to find the superior features. According to the result shown in Table \ref{featable}, the spectrum and MFCC perform the best among 1D and 2D features separately, with both achieving accuracies of over 95\%. The spectrum of ship-radiated noise consists of continuous, line, and modulation spectra, providing essential information about ship operations, engine status, and underwater environments. The reason why MFCC performs well may because it is designed based on human auditory perception and process signals through Mel filters. These characteristics better capture features at different frequencies in ship-radiated noise.

Based on the above experimental task results, in underwater target recognition task of identifying a specific type of ship signals from multiple targets, using spectrum and MFCC as feature inputs and DenseNet as classifier can achieve better recognition performance. Empirically, similar choices in practical applications have indeed yielded better classification results, consistent with our experimental findings.

\section{Conclusion}
\label{sec4}

\begin{table}[t!]\scriptsize
	\centering
	\caption{Avearge results of the same feature (\%)}
	{\fontsize{8}{10}\selectfont
		\label{featable}
		\centering
		\begin{tabular}{ccccccc}
			\Xhline{1pt}
			feature & accuracy & recall & precision & F1 score & FNR &FPR\\
			\Xhline{0.5pt}
			Spectrum & 96.73 & 96.43 & 96.30 & 96.35 & 3.49 & 3.65 \\
			MFCC & 95.47 & 94.92 & 95.06 & 94.95 & 5.36 & 4.54 \\
			Log Mel & 94.24 & 93.24 & 93.89 & 93.48 & 7.43 & 6.09 \\
			PSD & 93.73 & 92.40 & 93.55 & 92.87 & 7.84 & 7.35 \\
			GFCC & 92.77 & 91.43 & 92.25 & 91.75 & 10.23 & 6.66 \\
			PNCC & 92.56 & 91.25 & 91.96 & 91.54 & 10.16 & 7.25 \\
			DEMON & 89.61 & 85.46 & 87.62 & 86.29 & 17.02 & 11.92 \\
			LOFAR & 87.22 & 83.72 & 86.89 & 84.83 & 20.99 & 11.58 \\
			\Xhline{1pt}
		\end{tabular}%
	}
\end{table}

In this paper, we introduced QiandaoEar22, an underwater acoustic multi-target dataset constructed through the experiment collection in Qiandao Lake in 2022 and executed an experimental task to identify specific ships from the multiple targets to demonstrate the use of the dataset. We set speedboat, KaiYuan, and UUV as the research object and extracted different features and deep learning networks for classification. The best recognition accuracy of UUV target is 97.78\%, and we found using spectrum and MFCC as feature inputs and DenseNet as classifier can achieve excellent recognition performance. Our work established a benchmark for algorithm evaluation and may help the innovation development of UATD and UATR systems.
\bibliographystyle{IEEEtran}
\bibliography{IEEEabrv,mybibfile}

\end{document}